\documentclass[prd,superscriptaddress,showpacs,preprintnumbers,amsmath,amssymb]{revtex4}

\usepackage{graphicx}% Include figure files
\usepackage{dcolumn}% Align table columns on decimal point
\usepackage{bm}% bold math

\begin{document}

\title{Violation of Strong Energy Condition in Effective Loop Quantum Cosmology}
\author{Hua-Hui Xiong}
\email{jimhard@163.com}
\affiliation{Department of Physics, Beijing
Normal University, Beijing 100875, China}
\author{Jian-Yang Zhu}
\email{zhujy@bnu.edu.cn}
\affiliation{Department of Physics, Beijing
Normal University, Beijing 100875, China}
\date{\today}

\begin{abstract}
In loop quantum cosmology nonperturbative modification to a scalar matter
field at short scales implies inflation which also means a violation of the
strong energy condition. In the framework of effective Hamiltonian we
discuss the issue of violation of the strong energy condition in the
presence of quantum geometry potential. It shows that the appearance of
quantum geometry potential strengthens the violation of the strong energy
condition in small volume regions. In the small volume regions
superinflation can easily happen. Furthermore, when the evolution of the
universe approaches the bounce scale, this trend of violating the strong
energy condition can be greatly amplified.
\end{abstract}


\maketitle

\section{Introduction}

In classical general relativity the singularity theorems play an important
role in the quest of the property of spacetime evolution\cite{Hawking}.
These theorems predict that if the matter stress tensor satisfies the
so-called strong energy condition then the backward evolution of a globally
hyperbolic spacetime is singular. It means that the spacetime is
geodesically incomplete. Specialized to the context of isotropic and
homogeneous cosmology, the singularity theorem tells us that if the matter
satisfies the strong energy condition the scale factor (or the size of the
universe) will vanish in a finite time when the universe evolves backward.
This means that the universe begins from a singular point (corresponding to
the vanishing scale factor), and at the singular point matter density and
spacetime curvature diverge.

The finite expansion time of universe implies that there exists a particle
horizon, which is defined by the proper distance a particle traveled in the
past in the expanding universe. The particle horizon demarcates the causal
contact region in which events can connect each other by light signal. So
the presently observable universe is limited by the particle horizon in
principle. The particle horizon now is far bigger than that in the early
universe, but it does not correspond to the only one horizon in the early
universe when evolved backward. In fact, according to the standard big bang
model the horizon of the present universe is developed from many regions
which had no causal contact between them in the early universe. While
according to the observed data the cosmic microwave background (CMB)
radiation highly abides the Planck black body radiation spectrum. This shows
that the early universe in a large region is in a thermal equilibrium state.
The contradiction between the present homogeneous CMB radiation in a large
region and the apparently small horizon in the early universe is called the
horizon puzzle in cosmology. In the 1980s Guth proposed an inflation
mechanism to solve this problem\cite{Guth}. It was supposed that the early
universe experienced a short time inflation such that the particle horizon
increased quickly. In this inflation model the early universe is described
by a scalar field with self-interaction potential, while it seems unnatural
that in order to have sufficient amount of inflation with graceful exit the
potential must be fine-tuned. However, the inflation mechanism is still a
successful way to solve a series of problems in cosmology. Finally, it is
clear that the inflation model violates the strong energy condition.

We have known that when the scale factor tends to zero there exists a
singular spacetime point with diverging matter density. It is generally
accepted that the appearance of singularity in classical general relativity
means the failure of classical theory in the very small volume regions of
spacetime. So it is expected that in the small volume regions (Planck scale)
quantum gravity theory should replace the classical gravity and resolve the
singularity appearing in classical general relativity. So far, there are two
promising candidates for quantum gravity. One is string theory \cite{string}
and the other is loop quantum gravity (LQG)\cite{lqg}. LQG is background
independent and nonpertubative. The underlying geometry is discrete at
Planck scale. Loop quantum cosmology (LQC) inherits this feature by
introducing symmetry reduction on the level of quantum state (known as spin
network state)\cite{symmetryreduction}. Applying techniques developed in the
full theory (LQG) the classical Hamiltonian constraint which is reformulated
in terms of a new set of variables is quantized and the obtained quantum
Hamiltonian constraint can be interpreted as a quantum difference equation
which evolves non-singularly through the big bang point\cite
{lqc,mathematical-structure}. The quantum difference equation contains the
information as to the evolution of the universe at the Planck scale. However
the difference equation is difficult to solve. There are lacks of exact
solutions as well as physical inner product. All these prohibit us to know
about the semiclassical behavior of the quantum difference equation.
Approximated way is developed to tackle this issue. An exact coherent state
was constructed, but the explicit form of this coherent state depends on
numerical result of the quantum difference equation\cite{semiclassical-state}%
. In this paper our discussion is based on an effective Hamiltonian which
comes from the WKB approximation of the quantum difference equation\cite
{effective-Hamiltonian}.

LQC is free of singularity and the initial condition is determined by the
dynamical law (difference equation)\cite{singularity-free,initial-condition}%
. This happens at deep quantum region. However, it is necessary to know
about the property of the difference equation in semiclassical region. In
the semiclassical region it is assumed that spacetime geometry recovers its
continuous form and classical equation get modification from LQC. In this
region phenomena are investigated, such as a natural inflation from quantum
geometry\cite{inflation-geometry}, avoidance of a big crunch in closed
cosmology\cite{avoidance-big-crunch}, appearance of a cyclic universe\cite
{cyclic-universe} and a mass threshold of black hole\cite{mass-threshold},
etc. All these essentially captures the feature of the inverse scale factor
operator which greatly modifies the matter field at the scale $p\ll p_j$, $%
p_j=\frac 13\gamma \mu _0jl_p^2$, where $\gamma $ is the Barbero-Immirzi
parameter and $p_j$ distinguishes the different region of the modification
from inverse scale factor operator. For $p\gg p_j$ it is in the classical
region. There is a systematic way which takes an effective Hamiltonian to
investigate these phenomena carefully. The effective continuum Hamiltonian
is extracted from the difference equation and assumed to live on the
pseudo-Riemann manifold. This depends on two steps. Namely, a continuum
approximation was taken such that the fundamental difference is replaced by
a second-order differential equation (Wheeler-DeWitt equation) and then a
WKB rout is followed to obtain an effective Hamiltonian\cite
{effective-Hamiltonian}. The valid region of the effective Hamiltonian is
above a bounce scale which also sets the smallest scale for semiclassical
region\cite{generic-bounce}. Below the bounce scale the effective
Hamiltonian breaks down and is carried over by the quantum difference
equation. At large volume regions ($p\gg p_0$, $p_0$ demarcates the
different scale for effect of gravity) the effective Hamiltonian recovers
the classical form (classical Hamiltonian constraint).

In the framework of effective Hamiltonian the inflation and the bounce
happen independently of the initial condition and quantization ambiguity
parameter. These effects are natural results of the quantum geometry. It
shows that the inflationary cosmology avoids singularity by experiencing a
bounce. This implies that the strong energy condition (SEC) is violated in
the effective theory. However the violation of SEC does not only come from
the contribution of the modified matter field. In the effective theory there
is the ``quantum geometry potential'' which is suppressed in large scale ($%
p\gg p_0$), moreover, although the quantum geometry potential appears from
the WKB approximation it roots in the quantum difference equation and
predicts the existence of bounce as well as the bounce scale\cite
{generic-bounce}. Thereefore, it is necessary to investigate its role in the
evolution of the early universe. At the region $p_{bounce}<p<p_j$ ($%
p_{bounce}$ denotes the bounce scale) it is necessary to consider about the
effect of the quantum geometry potential, especially in the region
approaching the bounce scale where the quantum geometry potential greatly
changes the behavior of the effective Hamiltonian. In this paper we discuss
the issue of violation of SEC in flat cosmology ($\eta =0$) and mainly focus
on the role of the quantum geometry potential. The appearance of the quantum
geometry potential introduces a bounce which avoids the singular evolution
of the universe. So it can be expected that the existence of the quantum
geometry potential can help the violation of SEC. We shall show in this
paper that the appearance of quantum geometry potential will strengthen the
violation of SEC in small volume regions, furthermore, when the evolution of
the universe approaches the bounce scale, this trend of violating SEC can be
greatly amplified.

Based on the same slowly varying condition as in \cite{effective-Hamiltonian}%
, there is a direct way to obtain the effective Hamiltonian by using WKB
approximation which is referred as discrete correction to the effective
Hamiltonian\cite{discrete}. In small volume as noted in \cite{discrete} the
modified matter density and pressure go over to those defined in \cite
{effective-Hamiltonian}. In this paper our discussion of violation of SEC is
in small volume, so we do not consider about the discrete correction to the
effective Hamiltonian. Recently, different from the WKB way there is another
approximated method based on constructing the semiclassical state which
predicts the quadratic energy density modifications to the Friedmann equation%
\cite{quadratic}. However, in \cite{semiclassical-state} the analysis shows
that the WKB approximation matches well with the semiclassical state till
approaching the Planck scale. So, it is not expected that our analysis based
on the WKB way can be changed appreciably by the new approximated method.

This article is organized as follows. In Sec.\ref{Sec. 2}, we simply
introduce the process of obtaining the effective the effective Hamiltonian
and present the effective state parameter equation in the context of
effective Hamiltonian. Then in Sec.\ref{Sec. 3}, for a simple scalar field
we analyze the violation of SEC in the presence of the quantum geometry
potential. Finally, the Sec.\ref{Sec. 4} is the conclusion.

\section{SEC in the framework of effective Hamiltonian}

\label{Sec. 2}In LQC the Hamiltonian constraint is described by a difference
equation\cite{mathematical-structure}
\begin{equation}
0=A_{\mu +4\mu _0}\psi _{\mu +4\mu _0}-\left( 2+4\mu _0^2\gamma
^2\eta \right) A_\mu \psi _\mu +A_{\mu -4\mu _0}\psi _{\mu -4\mu
_0}+8\kappa \gamma ^2\mu _0^3\left( \frac 16\gamma l_p^2\right)
^{-1/2}H_m\left( \mu \right) \psi _\mu, \forall \mu \in R \label{b1}
\end{equation}
where $A_\mu :=\left| \mu +\mu _0\right| ^{3/2}-\left| \mu -\mu _0\right|
^{3/2}$, and $H_m\left( \mu \right) $ is the eigenvalue of matter
Hamiltonian which is assumed coupling with gravity via metric component and $%
\mu _0$ is a dimensionless parameter whose value is fixed by the length of
fiducial curve. Here, $\eta =0$ and $\eta =1$ correspond to flat and closed
models, respectively.

Because of the nonseparable structure of the kinematical Hilbert
space, there are infinite solutions for the Hamiltonian constraint.
In the absence of a physical inner product it is helpful to
stipulate the slowly varying property for a class of the solutions.
So the solution $\psi _\mu $ and coefficients in the difference
equation can be expressed as the function
of the continuous variable $p\left( \mu \right) $, $p\left( \mu \right) :=%
\frac 16\gamma l_p^2\mu $. In terms of slowly varying $\psi (p)$ the
difference equation can be approximated as a second order differential
equation
\begin{equation}
0=B_0\left( p,p_0\right) \psi \left( p\right) +4p_0B_{-}\left( p,p_0\right)
\psi ^{\prime }\left( p\right) +8p_0^2B_{+}\left( p,p_0\right) \psi ^{\prime
\prime }\left( p\right) ,p_0:=\frac 16\gamma l_p^2\mu _0,  \label{b0}
\end{equation}
where
\[
B_0\left( p,p_0\right) =A\left( p+4p_0\right) -\left( 2+144\frac{p_0^2}{l_p^4%
}\eta \right) A\left( p\right) +A\left( p-4p_0\right) +\left( 288\kappa
\frac{p_0^3}{l_p^4}\right) H_m\left( \mu \right) ,
\]
\[
B_{\pm }\left( p,p_0\right) =A\left( p+4p_0\right) \pm A\left( p-4p_0\right)
,
\]
and
\[
A\left( p\right) =\left| p+p_0\right| ^{3/2}-\left| p-p_0\right| ^{3/2}.
\]
And then the WKB way is used to get an effective Hamiltonian which is given
by
\begin{equation}
H^{eff}\left( p,K,\phi ,p_\phi \right) =-\frac 1\kappa \left[ \frac{%
B_{+}\left( p,p_0\right) }{4p_0}K^2+\eta \frac{A\left( p\right) }{2p_0}%
\right] +V_g+H_m\left( p,\phi ,p_\phi \right) ,  \label{b2}
\end{equation}
where the Poisson bracket between the extrinsic curvature $K$ and the triad
variable $p$ is $\frac \kappa 3$\cite{effective-Hamiltonian}. Here, $%
V_g=\left( \frac{l_p^4}{288\kappa p_0^3}\right) \left\{ B_{+}\left( p\right)
-2A\left( p\right) \right\} $. The quantum geometry potential is denoted as $%
V_g$ and is negative for $p>0$ (we only consider the positive value of $p$
because the negative $p$ corresponds to an inverse orientation universe).
For $p\gg p_0$, the classical Hamiltonian constraint is recovered. Next, we
are limited in the flat cosmology, i.e., $\eta =0$.

In terms of the effective Hamiltonian and comparing with the usual FRW
equation, the effective perfect fluid density and pressure are identified as
\begin{eqnarray}
\rho ^{eff} &=&\frac{32}3\frac \alpha {a^4}\left( H_m+V_g\right) ,
\label{b3} \\
P^{eff} &=&\frac{32}9\frac \alpha {a^4}\left\{ \left( 1-\frac{a\alpha
^{\prime }}\alpha \right) H_m-a\frac \partial {\partial a}H_m\right\} +\frac{%
32}9\frac \alpha {a^4}\left\{ \left( 1-\frac{a\alpha ^{\prime }}\alpha
\right) V_g-aV_g^{\prime }\right\}  \nonumber
\end{eqnarray}
where $\alpha =\frac{B_{+}\left( p\right) }{4p_0}$, $a$ is the FRW scale
factor and the relation between $p$ and $a$ is $p=\frac{a^2}4$\cite
{effective-Hamiltonian}. The prime denotes $\frac d{da}$.

In the context of isotropic and homogeneous cosmology the SEC becomes $4\pi
G(\rho +3P)\geqslant 0$, $\rho \geqslant 0$, where $\rho $ and $P$ are the
total energy density and pressure of the mater field. Usually SEC is
described by a state parameter equation which is defined as
\begin{equation}
\omega :=\frac P\rho .  \label{b4}
\end{equation}

In the framework of the effective Hamiltonian we still work in the
continuous spacetime as in classical theory, so SEC can be employed in this
effective theory. In terms of (\ref{b4}) the effective state parameter is
expressed as
\begin{equation}
\omega ^{eff}=\frac{P^{eff}}{\rho ^{eff}}=\frac 13\left( 1-\frac{a\alpha
^{^{\prime }}}\alpha \right) -\frac a3\frac{\frac \partial {\partial a}%
H_m+V_g^{\prime }}{H_m+V_g}  \label{b5}
\end{equation}

In the large volume regions, the quantum geometry potential $V_g$ is
suppressed and tends to vanish, so the effective state parameter equation
becomes
\begin{equation}
\omega ^{eff}\rightarrow \frac 13\left( 1-\frac{a\alpha ^{\prime }}\alpha
\right) -\frac a3\frac{\frac \partial {\partial a}H_m}{H_m}.  \label{b6}
\end{equation}
The violation of SEC has been carefully discussed in the absence of the
quantum geometry potential in \cite{generic-inflation} where in the small
volume regions for a positive scalar field potential $\omega
^{eff}\rightarrow -1$.

For the flat model ($\eta =0$), the effective Hamiltonian constraint behaves
as
\begin{equation}
H^{eff}=-\frac 1\kappa \frac{B_{+}\left( p,p_0\right) }{4p_0}K^2+V_g+H_m=0.
\label{b7}
\end{equation}
The above equation implies that
\begin{equation}
H_m+V_g=\frac 1\kappa \frac{B_{+}\left( p,p_0\right) }{4p_0}\geqslant 0.
\label{b8}
\end{equation}

In equation (\ref{b8}) the equality indicates an occurrence of bounce which
also suggests that there exists a smallest scale (bounce scale $p_{bounce}$)
and below this scale it is in a classically inaccessible region\cite
{generic-bounce}. At the bounce scale, $\rho ^{eff}=0$. So $p_{bounce}$ is a
singular point for the effective state parameter equation, in other words
the effective state parameter equation defines illegally at $p_{bounce}$.
However the Hamiltonian equation behaves well at bounce scale. The region
what we care about is above bounce scale. What is more, the effective
Hamiltonian (\ref{b7}) needs that matter field $H_m$ must be positive
definite because the kinetic term and the quantum geometry potential are
also negative value.

The occurrence of bounce (which also means $\rho ^{eff}\rightarrow 0$ when $%
p $ approaches $p_{bounce}$) essentially depends on the existence of the
quantum geometry potential. Next, we will show the violation of SEC for a
minimal coupled scalar field.

\section{Violation of SEC for a scalar field}

\label{Sec. 3}In LQC in the semiclassical region ($p_{bounce}<p<p_j$) a
classical scalar field gets modification. The modified scalar field is
obtained by replacing the $a^{-3}$ in the kinetic term by a function coming
from the definition of the inverse triad operator\cite
{quantization-ambiguity,bianchi-IX-model}. The modified scalar field with a
self-interaction potential is given by
\begin{equation}
H_m=\frac 12\left| \widetilde{F}_{j,l}(a)\right| ^{3/2}p_\phi ^2+a^3V(\phi ),
\label{b9}
\end{equation}
where $\widetilde{F}_{j,l}(a)=\left( \frac 13\gamma \mu _0jl_p^{2-1}\right)
F_l\left[ \left( \frac 13\gamma \mu _0jl_p^2\right) ^{-1}a^2\right] $ is a
smooth approximation (except at one point) of the inverse scale operator,
and
\begin{eqnarray}
F_l(q) &:&=\left( \frac 3{2(l+2)(l+1)l}\left\{ (l+1)\left[
(q+1)^{l+2}-\left| q-1\right| ^{l+2}\right] \right. \right.  \nonumber \\
&&\left. \left. -(l+2)q\left[ (q+1)^{l+1}-sgn(q-1)\left| q-1\right|
^{l+1}\right] \right\} \right) ^{\frac 1{1-l}}  \nonumber \\
&\rightarrow &\left\{
\begin{array}{c}
q^{-1},\ \ (q\gg 1), \\
\left( \frac{3q}{l+1}\right) ^{\frac 1{1-l}},\ \ (0<q\ll 1).
\end{array}
\right.  \label{b10}
\end{eqnarray}
The $j$ and $l$ are two quantization ambiguity parameters with $j$ being a
half integer and $l\in \left( 0,1\right) $. For large value of $j$ it can
lead to observable effects\cite{quantization-ambiguity}. For the modified
scalar field, the effective state parameter is
\begin{eqnarray}
\omega ^{eff} &=&\frac 13\left( 1-\frac{a\alpha ^{\prime }}\alpha \right) -%
\frac a3\frac{\frac 34\left[ \widetilde{F}_{j,l}(a)\right] ^{1/2}\left[
\widetilde{F}_{j,l}(a)\right] ^{\prime }p_\phi ^2+3a^2V\left( \phi \right) }{%
H_m+V_g}-\frac a3\frac{V_g^{\prime }}{H_m+V_g}  \nonumber \\
&=&\frac 13\left( 1-\frac{a\alpha ^{\prime }}\alpha \right) -\frac{\frac 12%
\left[ \widetilde{F}_{j,l}(a)\right] ^{3/2}p_\phi ^2\frac{q\frac d{dq}%
F_l\left( q\right) }{F_l\left( q\right) }+a^3V\left( \phi \right) }{H_m+V_g}-%
\frac a3\frac{V_g^{\prime }}{H_m+V_g},  \label{b11}
\end{eqnarray}
where $q=\left( \frac 13\gamma \mu _0jl_p^2\right) ^{-1}a^2.$

In the small volume regions, i.e., $0<q\ll 1$, the effective state parameter
equation is
\begin{equation}
\omega ^{eff}=\frac 13\left( 1-\frac{a\alpha ^{\prime }}\alpha \right) -%
\frac 1{1-l}+\frac{la^3V\left( \phi \right) }{\left( 1-l\right) \left(
H_m+V_g\right) }-\frac{-3V_g+\left( 1-l\right) aV_g^{\prime }}{3\left(
1-l\right) \left( H_m+V_g\right) }.  \label{b12}
\end{equation}

The difference equation (\ref{b1}) evolves forward with the fixed step $4p_0$
. When scale is below the fixed step it is in deep quantum range where the
continuous approximation becomes poor. So we can safely say that the region
(semiclassical region) what we consider about should be above the fixed
step. For convenience, we take $p\geqslant 5p_0$ for discussion below. But
the valid region of the effective Hamiltonian is $p\geqslant p_{bounce}$. As
for the relation between $p_{bounce}$ and the fixed step we will discuss in
the conclusion.

The first term in (\ref{b12}),
\begin{equation}
\frac 13\left( 1-\frac{a\alpha ^{\prime }}\alpha \right) \approx \left\{
\begin{array}{c}
-0.41,\ p=5p_0, \\
-2.7\times 10^{-4},\ p=100p_0, \\
0,\ p>100p_0.
\end{array}
\right. .  \label{c1}
\end{equation}
So the effect of the first term becomes negligible compared with the left
three terms.

The second term $\frac 1{1-l}$ in (\ref{b12}) is independent of the
particular form of the matter field and completely comes from the
modification of the inverse scale factor to the matter field in the small
volume regions. This depends on that in the small volume regions the inverse
scale factor is an increasing function with power $\frac 1{1-l}$ which
greatly differs from its classical behavior. The value of the second term is
only determined by the quantization ambiguity parameter $l$. It is clear
that $l\in \left( 0,1\right) $ makes $\frac 1{1-l}>1$. So this term causes $%
\omega ^{eff}<-1$ and implies the possibility of a superinflation. The final
result will rely on this term and the last two terms.

The last two terms in (\ref{b12}) correspond to the matter part and the
quantum gravity potential, respectively. From (\ref{b8}) we know that the
denominators of these two terms are all positive. Whether these two terms
strengthen or weaken a violation of SEC will depend on the signs of their
numerators. The third term indicates that for the violation of SEC the
potential of the scalar field behaves to differ from the view in \cite
{generic-inflation} where an appearance of a positive scalar potential
always leads to a violation of SEC and makes $\omega ^{eff}\rightarrow -1$
in small volume regions. However, here a negative scalar potential always
strengthens the violation of SEC. Conversely, a positive potential weakens
the violation.

The last term purely comes from the quantum geometry potential. In this term
the numerator is a monotonically decreasing function whose value is positive
and infinitely tends to zero for $p\gg p_0$, and
\begin{equation}
-3V_g+\left( 1-l\right) aV_g^{\prime }=\left\{
\begin{array}{c}
\lbrack 4.71+(1-l)12.73]\frac{p_0^{3/2}}\kappa ,\ p=5p_0 \\
\rightarrow 0,\ p\gg p_0
\end{array}
\right. .  \label{c2}
\end{equation}
Because $-\frac{-3V_g+\left( 1-l\right) aV_g^{\prime }}{3\left( 1-l\right)
\left( H_m+V_g\right) }<0$, the contribution of the quantum geometry
potential always strengthens the violation of SEC. This shows that the
appearance of the quantum geometry potential not only determines the
occurrence of a bounce, but also leads to an accelerated expansion of the
universe.

From the above analysis we know that in the small volume regions if $V\left(
\phi \right) <0$, then $\omega ^{eff}<-\frac 1{1-l}$. Since $\omega
^{eff}<-1 $ there exists a phase of superinflation. Compared to the result
in \cite{generic-inflation} a superinflation is more easily attained in this
context. In \cite{generic-inflation} a superinflation exists only for a
massless scalar field. Here, a negative or vanishing scalar potential can
lead to superinflation inevitably in the small regions. Because of the
existence of the quantum geometry potential, even for a positive scalar
potential
\begin{equation}
V\left( \phi \right) \leqslant \frac 13\frac 1{la^3}\left[ -3V_g+\left(
1-l\right) aV_g^{\prime }\right]  \label{c3}
\end{equation}
the effective state parameter equation is still $\omega ^{eff}<-\frac 1{1-l}$%
. Therefore, here a superinflation can happen in small volume regions only
if the scalar potential satisfies the condition (\ref{c3}). So the happening
of superinflation is independent of initial condition completely. Although
the quantization ambiguity parameter $l$ appears in the condition (\ref{c3}%
), $l$ only limits the upper bound of a scalar potential for occurrence of a
superinflation. The condition (\ref{c3}) shows that if $l\rightarrow 0^{+}$,
there must be a superinflation which is independent of the scalar potential.
If $l\rightarrow 1^{-}$, a superinfaltion can happen for $V\left( \phi
\right) \leqslant \frac{-V_g}{a^3}$ (the upper bound is a positive
potential.) and at the same time $\omega ^{eff}\ll -1$.

For the effective equation (\ref{b12}) there is an effect that when $p$
approaches the bounce scale the last two terms can be greatly amplified
because of the small value of their numerators (i.e., $H_m+V_g\sim 0$). For
the last term when $p\rightarrow p_{bounce}$, $\frac{-3V_g+\left( 1-l\right)
aV_g^{\prime }}{3\left( 1-l\right) \left( H_m+V_g\right) }\gg 1$. So, if $%
V\left( \phi \right) \leqslant 0$, $\omega ^{eff}\ll -1$ at the region $%
p\rightarrow p_{bounce}$.

There is another issue of the graceful exit. Only in the presence of the
modification to matter field from the inverse scale factor the inflationary
phase automatically ends when the peak of $F_l\left( q\right) $ is reached%
\cite{inflation-geometry}. However, from equation (\ref{b11}) it shows that
the existence of the quantum geometry potential can prolong the ``exit
time'', i.e., $p_{exit}>p_{peak}$, ( $p_{exit}$ denotes the value of $p$ at
the end of infaltion in the presence of the quantum geometry and $p_{peak}$
corresponds to the value of $p$ at the peak of $F_l\left( q\right) $),
because at the right side of the equation (\ref{b11}) the third term is
independent of the modification from the inverse scale factor and it is
always a negative value. But still a graceful exit can happen for some value
of $p$ because when $p$ increases the quantum geometry potential tends to
vanish.

Now let us concern the region discussed in this paper. As in \cite
{effective-Hamiltonian,generic-bounce,generic-inflation}, we work in the
valid regions of the effective Hamiltonian which also corresponds to the
semiclassical regions, i.e., $p_0<p_{bounce}<p<p_j$. Here $p_0$ is the
quantum geometry scale and $p_j$ is the inverse scale factor scale. When $%
p\gg p_j$ it is in the classical regions.

\section{Conclusion}

\label{Sec. 4}In this paper we mainly discuss the issue of the violation of
SEC in the presence of the quantum geometry potential. It shows that the
appearance of the quantum geometry potential strengthens the trend of the
violation in the small volume regions, especially in the region next to the
bounce scale. A superinfaltion can happen if the scalar potential satisfies
the given condition (\ref{c3}). In conclusion, the inflation from quantum
geometry is raised by two parts in which one is the modified matter and the
other is the quantum geometry potential. In the small volume regions they
are all important for the violation of SEC. As in usual cosmology described
by the classical general relativity a negative pressure can lead to an
accelerated expansion of the universe. Similarly, in effect the quantum
geometry potential behaves like a negative pressure for the modified FRW
equation by LQC.

One open issue is the range of the semiclassical region. In the
semiclassical region $p_{bounce}$ and $p_j$ are two basic scales. In \cite
{semiclassical-state} the numerical result shows that the effective theory
based on the WKB approximation can be trusted to approach the Planck scale $%
l_p$. So the bounce scale is comparable with $p_0$. As for $p_j$, it is an
open problem to determine the value of $j$. But it is expected that $j$ is
large enough to ensure that the semiclassical region is big enough in order
that the phenomena can leave observable effects\cite{CMB-spectrum}.
Therefore, our discussion in this paper essentially assumes that $p_j\gg p_0$%
.

In this paper we mainly care about the effect of the quantum geometry
potential for the violation of SEC. However, because the quantum geometry
potential appears in the effective Hamiltonian, basing on the effective
Hamiltonian, many phenomena can be investigated again. And, it is expected
that in the presence of the quantum geometry potential the results obtained
in the before can be changed quantitatively or qualitatively in the small
volume regions.

\acknowledgments
The work was supported by the National Basic Research Program of China
(2003CB716302).

\end{document}